\documentstyle{article}
\begin{document}

\begin{flushright}
                    KEK CP-032~~~~~~~~\\
                    KEK preprint 95-51\\
                    ENSLAPP-A-522/95~~\\
                    LPTHE-Orsay 95/37
\end{flushright}
\renewcommand{\baselinestretch}{1.5}
\large

\begin{center}
{\Large {\bf A Parton Shower Model for \\
Hadronic Two-Photon Process in $e^+e^-$ Scatterings} }
\vskip 0.8in
   T. MUNEHISA
\vskip 0.2in
Faculty of Engineering, Yamanashi University
\vskip 0.1in
Takeda, Kofu, Yamanashi 400, Japan
\vskip 0.1in
     P. AURENCHE,~~~  J.-Ph. GUILLET
\vskip 0.2in
Laboratoire de Physique Th{\' e}orique ENSLAPP$-$ Groupe d'Annecy
\vskip 0.1in
LAPP, IN2P3-CNRS, B.P.110, F-74941 Annecy-le-Vieux Cedex, France
\vskip 0.2in
     M. FONTANNAZ
\vskip 0.2in
Laboratoire de Physique Th{\'e}orique et Hautes Energies
\vskip 0.1in
Universit{\'e} de Paris XI, batiment 211, F-91405 Orsay Cedex, France
\vskip 0.2in
       Y. SHIMIZU
\vskip 0.2in
National Laboratory for High Energy Physics(KEK)
\vskip 0.1in
Oho 1-1 Tsukuba, Ibaraki 305, Japan
\eject

\end{center}
\vskip 0.5in
\begin{center}
{\bf ABSTRACT}
\end{center}
 A new model of QCD parton shower is proposed which is dedicated to
 two-photon process in $e^+ e^-$ scattering. When hadron jets are
produced, the photon may resolve into quark-antiquark pairs so that
the structure functions of the photon should be introduced.
Based on the Altarelli-Parisi equation for these functions,
an algorithm is formulated that allows us to construct a model
for parton showers for the photon.
Our model consists of two parts, one of which describes the
deep inelastic scattering of the photon and the other one the
scattering of two quasi-real photons. Using the model
some results are presented on parton distributions and jet production.

\eject

\def\GeV{\hbox{GeV}}

\noindent
{\bf Section 1. Introduction }

Recent extensive studies in TRISTAN experiments on the two-photon
process in $e^+e^-$ scattering, particularly on jet
production\cite{2ph1,2ph2}, have stimulated more detailed theoretical
investigations in QCD\cite{th,kin,QCD}.
However these calculations have been limited to the inclusive case
so far.
In order to make a precise comparison between experiments and theory,
one needs an event generator where one starts with the generation of
partons and then convert them into hadrons and other stable particles.
Distributions of generated partons are completely determined by
perturbative QCD. To generate partons one has two possibilities.
The first one is the matrix element method which uses the fixed order
perturbative calculation and the
other one is the parton shower method\cite{ll}  based on
the Altarelli-Parisi(AP) equation. In the latter one can take into
account all the leading contributions from the collinear singularities
without limitation in the order of the coupling.

The parton shower model has been demonstrated to be powerful and
indispensable for
quantitative studies in $e^+e^-$ annihilation and other processes.
It is also true for the two-photon experiments and it is highly
desirable to develop an event generator that includes parton showers,
since recent experiments provide data with rather good accuracy.
In this paper we construct such a parton shower model dedicated to
the two-photon process.

One of the main features of this parton shower is the resolution of the
photon, i.e. the case where the photon splits into a quark-antiquark
pair at the beginning of the reaction.
Partons inside the photon have two sources.
One is taken into account by the so called vector meson dominance
model(VDM) and the other one is
the perturbative photon. They correspond to the solutions of
the homogeneous and inhomogeneous AP equations, respectively.
In terms of parton language, the VDM contribution arises when
 the photon
converts into the vector mesons like $\rho^0$ and $\omega$ and then
the partons inside these mesons undergo QCD evolution.
On the other hand the perturbative photons are those that decay into
quark-antiquark pairs and then these quarks evolve according to perturbative
QCD.  We will call them the VDM and the perturbative photon parts in
the following.

In the case of two quasi-real photon scattering we have another
contribution where the original photon undergoes a hard scattering.
This shall be called the direct part.

The present model is limited to the leading-logarithmic(LL)
approximation. An extension to the next-to-leading logarithmic
case is, however, not difficult\cite{nll}.
One reason for this limitation is that experimental data are not yet
precise enough to determine the QCD scale parameter $\Lambda$ and
another
reason is that the most important contributions of the higher order
are already included into the model as discussed in\cite{nll}.

To apply the model to actual reactions
one has to distinguish two situations(see Fig.1).
The first one is the deep inelastic scattering of a photon,
i.e. one photon is quasi-real while the other has a large virtual
mass squared $-Q^2$. One requirement from perturbative QCD is that
this virtual mass squared is much larger than the QCD scale
parameter and hadron mass scale. Hence one has to assume
$Q^2 \geq 5-10 \ \hbox{\GeV}^2 $.
The second case is for two quasi-real photon scattering.
In this process one can get reliable predictions from QCD calculations
only when jets with large $p_T^2$ are produced. Notice that our event
generator produces only partons, not hadrons. Hence the VDM
contributions express only the part related to the hard scattering.
We do not consider the low $p_T$ scattering in the hadron interaction.

In the next section we show how to formulate the parton shower model,
starting from
the AP equations. In section 3 a detailed description of the model will
be presented. There we discuss the differences between
 the models of
deep inelastic scattering and two quasi-real photon scattering.
These are related not to the algorithm of the model but the way to
apply it.
Section 4 shows some results obtained by Monte Carlo simulations of
the parton shower. One will see that it reproduces the same
distributions as the conventional QCD calculations for inclusive
observables.  The last section is devoted to summary and discussion.

\vskip 0.5in
\noindent
{\bf Section 2. Basic formulation }

The algorithm of the parton shower is based on the AP
equation, which governs the evolution of
the photon structure function.
First we solve this equation in the way most convenient to our purpose,
which gives us a clear basis for the construction of the shower model.
For simplicity we limit the discussion here to the non-singlet case
within the LL approximation.

The AP equation for $q_{NS}(x,Q^2)$, the non-singlet quark
distribution in the photon\cite{th,kin}, is given by
\begin{equation}
  {d q_{NS}(x,Q^2) \over d \ln Q^2} = {\alpha_s\over 2\pi}
\int_x^1 {dy \over y} P_{qq}^{(0)}(x/y) q_{NS}(y,Q^2)
 +{\alpha\over 2\pi} k_{NS}^{(0)}(x), \label{eq:AP}
\end{equation}
Here $P_{qq}^{(0)}(x)$ is the quark splitting function.
This equation differs from the usual one
by the existence of an inhomogeneous term, $ k_{NS}^{(0)}(x)$.
It is known that this term has a $x$-dependence of
the form $x^2+(1-x)^2$.
The QED coupling is denoted as $\alpha$ and that of QCD is given by
\begin{equation}
 \alpha_s(Q^2)={4\pi \over \beta_0 \ln Q^2/\Lambda^2}
 \equiv {\alpha_0 \over \ln Q^2/\Lambda^2 },
 \ \  \ \ \beta_0 = 11-{2\over 3}N_F .
\end{equation}
where $N_F$ is the number of  flavors.
Eq.(1) can be solved easily if one takes its moments:
\begin{equation}
 q_{NS}(n,Q^2)= \int_0^1 dx x^{n-1}  q_{NS}(x,Q^2) ,
\end{equation}
\begin{eqnarray}
 q_{NS}(n,Q^2) &=& q_{NS}(n,Q^2_0) \left
({\alpha_s(Q^2_0)\over \alpha_s(Q^2)}\right)^{\alpha_0d(n)}
\nonumber  \\
  &+& {\alpha \over \alpha_s(Q^2)}  {a_{NS}(n)\over \alpha_0d(n)-1}
\left
\lbrack \left({\alpha_s(Q^2_0)\over \alpha_s(Q^2)}\right)^{\alpha_0d(n)
-1}-1 \right\rbrack,
\label{eq:sol1}
\end{eqnarray}
where $d(n)$ is the moment of $P_{qq}^{(0)}(x)/2\pi$ and $a_{NS}(n) $
is that of $(\alpha/2\pi) k_{NS}^{(0)}(x)$.

In order to show the dependence on $n$ in a more transparent form,
we introduce an integral to rewrite the second term,
\begin{equation}
q_{NS}(n,Q^2) =q_{NS}(n,Q^2_0) e^{\alpha_0d(n)\overline{s}}
+ {\alpha \over \alpha_s(Q^2)}a_{NS}(n)\int^{\overline{s}}_0
d\eta e^{-\eta} e^{\alpha_0d(n)\eta },
\end{equation}
where
\begin{equation}
  \overline{s}=\ln {\alpha_s(Q^2_0)\over \alpha_s(Q^2)}
= \ln {\ln(Q^2/\Lambda^2) \over \ln (Q^2_0/\Lambda^2)}.
\nonumber
\end{equation}
This is the final form of the solution in the moment space.

To get the corresponding solution in the $x$-space, we make the inverse
Mellin transformation defined by the formula
\begin{equation}
 f(x) = \int_{r_0-i \infty}^{r_0+i \infty} {dn \over 2\pi i}
 x^{-n} f(n),
\nonumber
\end{equation}
where $r_0$ is a real number which defines the location of
the integration path.
The result can be cast into the following form
\begin{eqnarray}
q_{NS}(x,Q^2) & = &  \int_x^1 {d y \over y} K_{NS}^{(0)}
(x/y,\overline{s})q_{NS}(y,Q^2_0)
\nonumber \\
 & +& \ln (Q^2/\Lambda^2) \int^{\overline{s}}_0
d  \eta e^{-\eta}  \int_x^1 {d y \over y} K_{NS}^{(0)}(x/y,\eta)
{\alpha \over 2\pi } k_{NS}^{(0)}(y).
\label{eq:sol}
\end{eqnarray}
Here we have introduced the QCD kernel function
$ K_{NS}^{(0)}(x,\overline{s}) $ defined in \cite{ksy}.
\begin{equation}
 K_{NS}^{(0)}(x/y,\overline{s})=\int_{r_0-i \infty}^{r_0+i \infty}
{dn \over 2\pi i} x^{-n} e^{\alpha_0 d(n)\overline{s}}.
\end{equation}

\noindent
This kernel function is parametrized in a simple form so that
it is easy
to carry out convolution integrals in Eq.(\ref{eq:sol}) numerically.
The physics meaning of each
term in this solution is clear. The first term expresses the
contribution from VDM, while the second contains
the effect of the perturbative photon.

In order to translate the Eq.(\ref{eq:sol}) into the language of
parton showers, it is convenient to change the integral variable $\eta$
as follows,
\begin{equation}
   \eta =\ln {\ln Q^2/\Lambda^2 \over \ln K^2/\Lambda^2 },
\end{equation}
where $-K^2$ can be regarded as the squared virtual mass of the quark
that is emitted from the photon. By this change of variable,
one can rewrite the solution in the following way:
\begin{eqnarray}
q_{NS}(x,Q^2) & = &  \int_x^1 {d y \over y} K_{NS}^{(0)}
(x/y,\overline{s})q_{NS}(y,Q^2_0)
\nonumber \\
  &+& \int^{Q^2}_{Q_0^2}
  {d K^2 \over K^2} \int_x^1 {d y \over y} K_{NS}^{(0)}(x/y,\eta(K^2))
{\alpha \over 2\pi } k_{NS}^{(0)}(y).
\label{eq:res}
\end{eqnarray}
The second term of this equation says that at the beginning the photon
resolves into a quark-antiquark pair and then the spacelike
quark(antiquark) evolves according to the usual QCD branching,
because $K_{NS}^{(0)}(x/y,\eta(K^2))$ expresses the evolution from
$K^2$ to $Q^2$. Based on this observation we can formulate an algorithm
for the QCD evolution of the photon. So far the discussion is limited
to the non-singlet case, but the extension to the singlet case can be
done in a parallel way and our model is, needless to say, constructed
to contain both contributions.

\vskip 0.5in
\noindent
{\bf  Section 3. The model}

In this section we describe the model in some details.
We assume the forward evolution method for the
spacelike QCD evolution, as there is no essential difference
between the forward evolution and the backward one\cite{tanaka1}. One
reason is, however, that the former makes it easier to write programs
according to Eq.(\ref{eq:AP}).

The photon branching differs slightly from the other processes such as
$e^+e^-$ annihilation or deep inelastic $ep$ scattering.
In the latter the total energy is conserved and
the anomalous dimension corresponding to it vanishes identically.
In the shower model which has been developed by one of the authors,
the algorithm
crucially relies on this conservation\cite{tanaka2}. In other words
the evolution is considered {\sl not} for the particle distributions
$q(x,Q^2)$, {\sl but} for the energy distributions $xq(x,Q^2)$.
In the present case, however, Eq.(\ref{eq:AP}) is inhomogeneous and
the second moment of $k_{NS}^{(0)}(y)$ does not
vanish so that the energy sum rule is broken. Hence in the photon
shower model, one has to abandon parton energy conservation.
Of course the total energy including that of the photon should
be conserved.
\begin{eqnarray}
 \int_0^1 dx x \lbrack \sum_f
( q_f(x,Q^2)+\overline{q}_f(x,Q^2))+ G(x,Q^2) \rbrack
\nonumber
\\
  = \int^{Q^2}_{Q_0^2} {d K^2 \over K^2}
{\alpha \over 2\pi }\int_0^1 d x x  k_{NS}^{(0)}(x)
\label{eq:eng}
\end{eqnarray}
Therefore the quantity appearing on the right-hand side of this
equation is taken into account as {\sl the weight of an event}.

Let us state the algorithm step by step. It can be divided
into the spacelike evolution and the hard scatterings that take place
afterwards.  Procedures of the spacelike evolution can be summarized as
follows;
\vskip 0.1in
(A) Choose  $Q^2$.

(B) Calculate the energy in the VDM, which is independent of $Q^2$, and
the energy in the perturbative photon by using Eq.(\ref{eq:eng}).
The sum of these energies is kept as the weight of an event.

(C) Determine the process, either the VDM or the perturbative photon,
    according to the ratio of energies.

(D) If the VDM is chosen, make the usual QCD evolution up to $Q^2$
    from $Q^2_0$, the cut-off momentum.

(E) In the case of the perturbative photon, determine the virtual mass
squared $K^2$ according to the probability $d K^2/K^2$.
Then choose a flavor of quark according to the ratio of charges squared.

(F) For the selected flavor of quark or antiquark, make the usual
QCD evolution up to $Q^2$ from $K^2$.
\vskip 0.1in

The same algorithm is used for the deep inelastic photon scattering and
two quasi-real photon scattering with large $p_T^2$. The hard
scattering parts, however, completely differ from each other.

First we discuss the deep inelastic case. Here the QCD evolution
gives the structure function $F_2(x,Q^2)$ of the photon, which is
connected to the cross section in $e\hbox{-}\gamma$ scattering through
the following relation\cite{kin},
$$  {d \sigma \over d x dQ^2 }= { 4\pi \alpha^2\over x Q^4}
  (1-y) F_2(x,Q^2),$$
$$  y=Q^2/(xs_\gamma) ,$$
\noindent
where $s_\gamma$ is the energy squared of the $e\hbox{-}\gamma$ system.

In the hard scattering of this process the gluon does not participate
but the quarks do with the weight of the squared charge.
Determination of the four momenta is completed after
the timelike evolution is developed and the energy of the initial
photon is fixed by the spectrum function of the photon in the
electron, which is given by the Weizs\"{a}cker-Williams approximation.
One of its simplest form is given by
$$  F_{\gamma/e}(z,E)={\alpha\over \pi z}
 (1+(1-z)^2)(\ln{E\over m_e} -{1\over 2}). $$
In the analysis of actual data more sophisticated versions are used.

Next we discuss the scattering of two quasi-real photons.
In this case contributions from the direct photon should be
taken into account.
The magnitude of the cross sections for various hard processes is
necessary to determine the weight of events. They are different
depending on the type of partons involved in the hard scattering.
For example, if the initial state of the hard scattering is a gluon
and a direct photon, the final state is a quark and an antiquark.
Then we have to know the cross section.
\begin{eqnarray}
 {d \sigma(\hat{s}) \over
d p_T^2}(G + \gamma \rightarrow q + \overline{q}) .
\noindent
\label{eq:cross}
\end{eqnarray}
Here $\hat{s}$ is the energy squared of this hard scattering and
the hard scale $Q^2$ can be identified as the transverse momentum
squared of the scattered partons, $Q^2=p_T^2$.

As the possible initial states of the hard scatterings we have to
consider $q\hbox{-}q$, $q\hbox{-}\overline{q}$, $q\hbox{-}G$,
$G\hbox{-}G$, $q\hbox{-}\gamma$, $G\hbox{-}\gamma$ and
$\gamma\hbox{-}\gamma$.

In order to generate events in the most effective way with $O(1)$
weights, we adopt the following method. The cross sections for hard
scattering with photons of direct, direct-resolved, doubly-resolved
are in the order of $\alpha^2$, $\alpha\alpha_s$ and $\alpha_s^2$,
respectively. Consequently the corresponding events must have quite
different weights. In the total cross section, however, the order of
$\alpha$ is common for these three channels. In other words the
probabilities that a single real photon resolves or not is
$O(\alpha):O(1)$. Hence we can rearrange $\alpha$ by removing it from
the hard cross section to the probability of the photon branching.
Thus we have the ratio $O(\alpha):O(\alpha)$ to branch or not for
a photon and $O(1):O(\alpha_s):O(\alpha_s^2)$ for the hard cross
sections, which are comparable in magnitude. This rearrangement then
allows us to generate events with good efficiency. We use
 $d \sigma^0 / d p_T^2 = \pi/p_T^4$
as the reference cross section. If any hard cross section is equal to
this, the event is generated with weight = 1.

We like to point out also that the argument of $\alpha_s$
in the hard cross section is $ p_T^2 $, not $\hat{s}$ in our model(in
contrast to the case of QCD branching where the virtuality
$K^2$ enters into the argument).
This is because there appear some $t$-channel diagrams in the hard
scattering, and when $p_T^2$ is smaller than $\hat{s}$, the cross
section is dominated by the $t$-channel contribution.
The running effect comes from the virtual corrections
in the $t$-channel propagator and its scale is developed to $p_T^2$.

When the final states of the hard scattering is fixed, we make timelike
evolutions from $Q^2$ for timelike partons. Then we fix the energy of
the initial photons in the electron and the positron, and thus
we can determine the four-momenta of all partons.

Finally a comment is given about the higher order
contributions. In the literature it has been emphasized that they are
important for quantitative analysis\cite{phnll,AGF}. It has been known,
however, from the study of $e^+e^-$ annihilation, that
the most important
contributions in the next-to-leading logarithmic terms can be included
in the LL shower by using the transverse momentum squared $p_T^2$ as
the argument of the QCD coupling $\alpha_s$ and imposing the angular
ordering\cite{nll}. This is also valid for the photon evolution we are
considering. These effects are easily implemented into our model.

\vskip 0.5in
\noindent
{\bf Section 4. Results}

 We present results on parton distributions.
These distributions are usually obtained by integrating the AP
equation numerically. The parton shower provides us an another method
to get them. Though this equivalence has been well established, it will
be instructive to understand the algorithm, by comparing the results
of the parton shower with those obtained by the conventional method.
Many kinds of distributions\cite{dist} have been proposed in the
literature.  First we use those given in ref. \cite{GRV} for the
LL approximation, because their distributions are parametrized in
simple functions.
In order to make precise comparison, we calculate parton
distributions by the shower model with a fixed cutoff $\epsilon $
to regularize the infrared singularity.
Results are shown in Fig.2, where one can see good
agreement which shows that the shower algorithm  yields the same
distribution as obtained by usual methods to solve the  AP
equation directly.

Next we discuss the higher order effects. As mentioned in section 3,
the most
important effect beyond the leading order can be easily included
into the LL model
by using  the transverse momentum squared $p_T^2$ as the argument
of  the running coupling.
To show it, we calculate   the structure function, $F_2(x,Q^2)$
by the parton shower with the transverse momentum $p_T^2$  and the
virtual mass $K^2$ as the arguments.
As the full next-to-leading logarithm(NLL) solution we take
the distributions given in \cite{AGF}.
Here one should note that  this comparison is not trivial. One reason
is that the parton distributions in the NLL order
are scheme-dependent and the definition of the initial  distributions
is not unique. Another reason is that in ref.\cite{AGF} the
evolution started at too small $Q^2_0$ so that contributions beyond
the NLL order cannot be ignored.
Therefore for the comparison we start the shower evolution at
$Q^2 = 4\GeV^2$ using $F_2(x,4\GeV^2) $ as
the initial parton distribution.
 From Fig.3 we find that our shower model with the running coupling of
$p_T^2$ can reproduce the NLL results qualitatively. This agreement
is more remarkable for higher $Q^2$.
However considering the fact that we are able to get a qualitative
agreement between the NLL calculation and the LL shower together with
the fact that there still remains non negligible experimenatal
errors in the present data, there seems no need for
the NLL shower model for the moment, though it is in principle
possible\cite{nll}.

Next we will discuss jet production in two quasi-real photon processes
in $ e^+ e^- $  scattering.
In order to check the program, we compare the results of our model
with the inclusive QCD calculations obtained by  integrating
the parton distributions $f_i(x,p_T^2)$ with the hard cross sections
$\sigma^H_{ij}$, where $i,j$ denote photon, quarks and gluon.
\begin{eqnarray}
 {d \sigma \over d p_T^2} & = &
\int_0^1 d z_1 F_{\gamma/e}(z_1,E)
\int_0^1 d z_2 F_{\gamma/e}(z_2,E)  \nonumber \\
 & & \times  \sum_{i,j}
 \int_0^1 dx_1 f_i(x_1,p_T^2) \int_0^1 dx_2 f_j(x_2,p_T^2)
\delta(z_1z_2x_1x_2 s - \hat{s}) {d \sigma^H_{ij}(\hat{s})
\over d p_T^2}
\end{eqnarray}

It should be noted that it is {\sl not} trivial for our
model to get the cross section,
because the parton distributions are obtained as a result of the
generated events\cite{QED}, not being assumed at the beginning of the
calculation. Figure 4, where the integrated cross sections,
 $ \int_{p^2_{T\ min}}^{s/4} d p^2_T d \sigma /d p^2_T $ are given,
 shows consistency of our model.
Here to make a detailed comparison, we present the results for
hard scattering of quark-photon and gluon-photon.
The contributions from other processes are small(parton-parton
scattering) in TRISTAN energy range, or are trivial(photon-photon
scattering).

The parton shower model has an advantage, when one wants to make
topological analyses of the final particles. As a typical example of
these kinds of analyses, we take the thrust distribution.
Since the system of final partons is boosted along $z$-direction,
we analyze events using transverse momentum.
Similarly to the case in $e^+e^-$ annihilation, the thrust distribution
changes as the energy increases(see Fig.5). Note again that in
the present analysis we make no hadronization and the results are
not fully meaningful as a physical prediction. The hadronization
should be included for quantitative analysis of actual data.

\vskip 0.5in
\noindent
{\bf Section 5. Summary and discussions}

In this paper we proposed an algorithm of parton shower model for
two-photon processes in $e^+ e^-$ scattering and constructed a
Monte Carlo model according to it.
We formulated the algorithm using AP equations and the kernel
function method\cite{ksy} for the QCD evolution.
In the construction of the model, we adopted the forward spacelike
evolution for the energy distributions of partons $xq(x,Q^2)$ by taking
account of the breaking of energy conservation due to
the inhomogeneous term in the AP equations for the photon.

Using this Monte Carlo model, we have shown the parton distributions
of the photon and some results on jet production in the two-photon
process. These results clearly demonstrated that the model is
consistent with perturbative QCD.

Our model will be completed as an event generator if combined with
some hadronization models. Then it can be used for reliable
estimation of the detector efficiency in experimental analysis.
On the other hand the model will be useful for theoretical
investigations, as the model is rigorously based on perturbative
QCD. For example, it is easy to change the input parton distributions
at $Q^2_0$ in our model. Thus we can study in detail how the change
of the input distributions affects the final results.
Also interesting is the study on charm
distributions. However, the value of $Q_{0c}^2$, where the photon
starts to resolve into charm quark pair, is not clear. Naively one
expects $4m_c^2$, but kinematically it should be $4m_c^2/x_b$,
where $x_b$ is a Bjorken variable.
The parton shower model presented in this  work
would give us a quantitative answer to the question whether
this kinematical change modifies the parton distributions or not.
Extensive studies on these points will be presented in a
forthcoming paper.

\vskip 0.5in
{\bf Acknowlegdements}

We would like to thank D. Perret-Gallix and M. Werlen for their
interest in our work and encouragement. We are indebted to our
colleagues of KEK working group(Minami-Tateya) for their help.
This work has been done under the collaboration of KEK and LAPP
supported by Monbusho, Japan and CNRS/IN2P3, France.
\noindent


\eject

\eject

{\bf Figure Captions}

Fig.1~~ The schematics of deep inelastic scattering of photon(a)
and two quasi-real photons process(b) in $e^+e^-$ scattering.
Here $\gamma$ denotes a quasi-real photon, while $\gamma^*$ denotes
the virtual one.

Fig.2~~ Parton distributions $xf(x,Q^2)/\alpha $.
 The solid and dashed curves are calculated by the parametrized form
given in ref.[14] for LL approximation. The circles and diamonds
show results of our shower model with
a constant cutoff, which should be very small compared to the
observed $(1- x)$ value. Here it is $10^{-4}$.
 The evolution starts from $Q^2=0.25\GeV^2$ by
taking common initial distributions. (a) $u$-quark distributions.
(b) $u$-quark and gluon distributions.

Fig.3~~   The structure function $F_2(x,Q^2)/\alpha$.
 The  curves are  calculated by the authors of ref. [12]
The diamonds are results by the shower model with $p_T^2$,
while the plus symbols represent that with the virtual mass squared.
These distributions are normalized at
$Q^2=4\GeV^2$. The full circles are the experimental
data[1]. $Q^2 = 73\GeV^2$(a), $ 1000\GeV^2$(b).

Fig.4~~ Cross sections of jet production integrated over the transverse
momentum in the two quasi-real photon process in $e^+ e^-$ scattering.
 The solid and dashed curves are calculated with the
parton distributions of ref.[14]. Circles and diamonds
 are results of our parton generator.
In the solid curve and circles
 the hard scattering contains only the photon-quark scattering,
while in the dashed curve and diamonds it contains only photon-gluon
scattering.
The c.m.s energy is $58\GeV$.

Fig.5~~ Thrust distributions of the final partons
in the two quasi-real photon process in $e^+ e^-$ scattering.
 This is calculated only by
transverse momenta of the final partons.
 The histogram, circles and diamonds are results
with $p_{Tmin}^2 = 20\GeV^2,\ 50\GeV^2 $ and $100\GeV^2$, respectively.
The c.m.s energy is $58\GeV$.

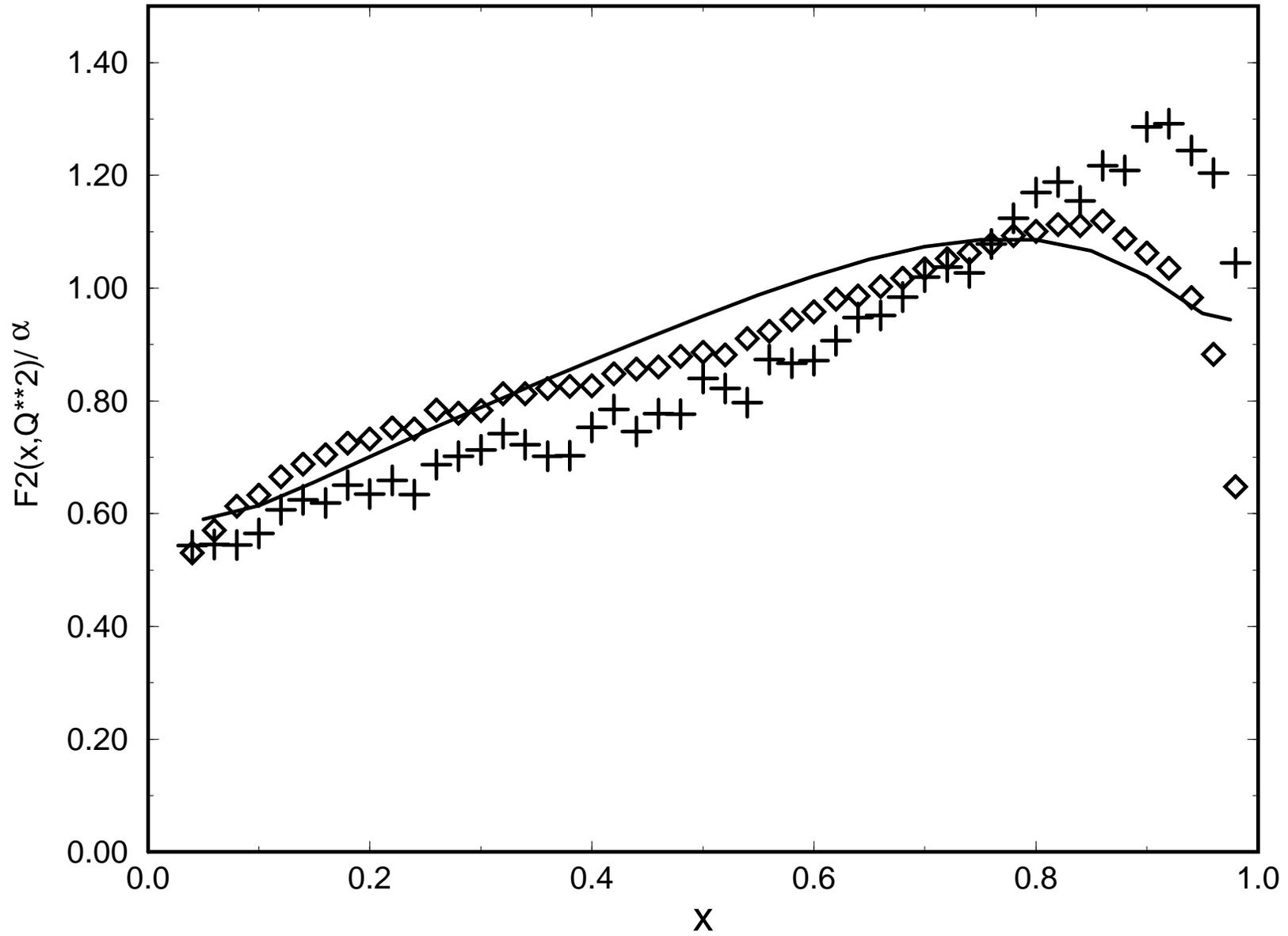
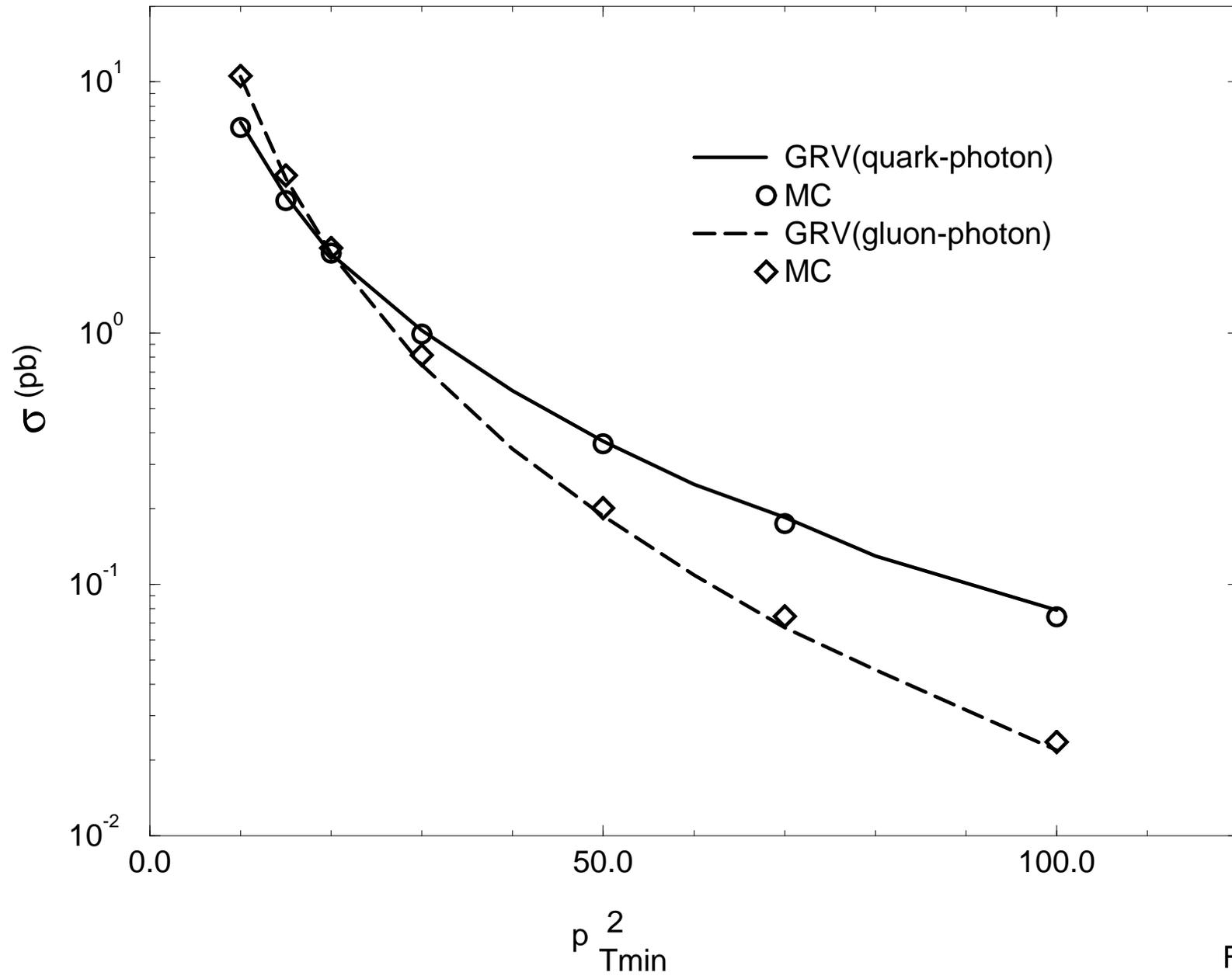
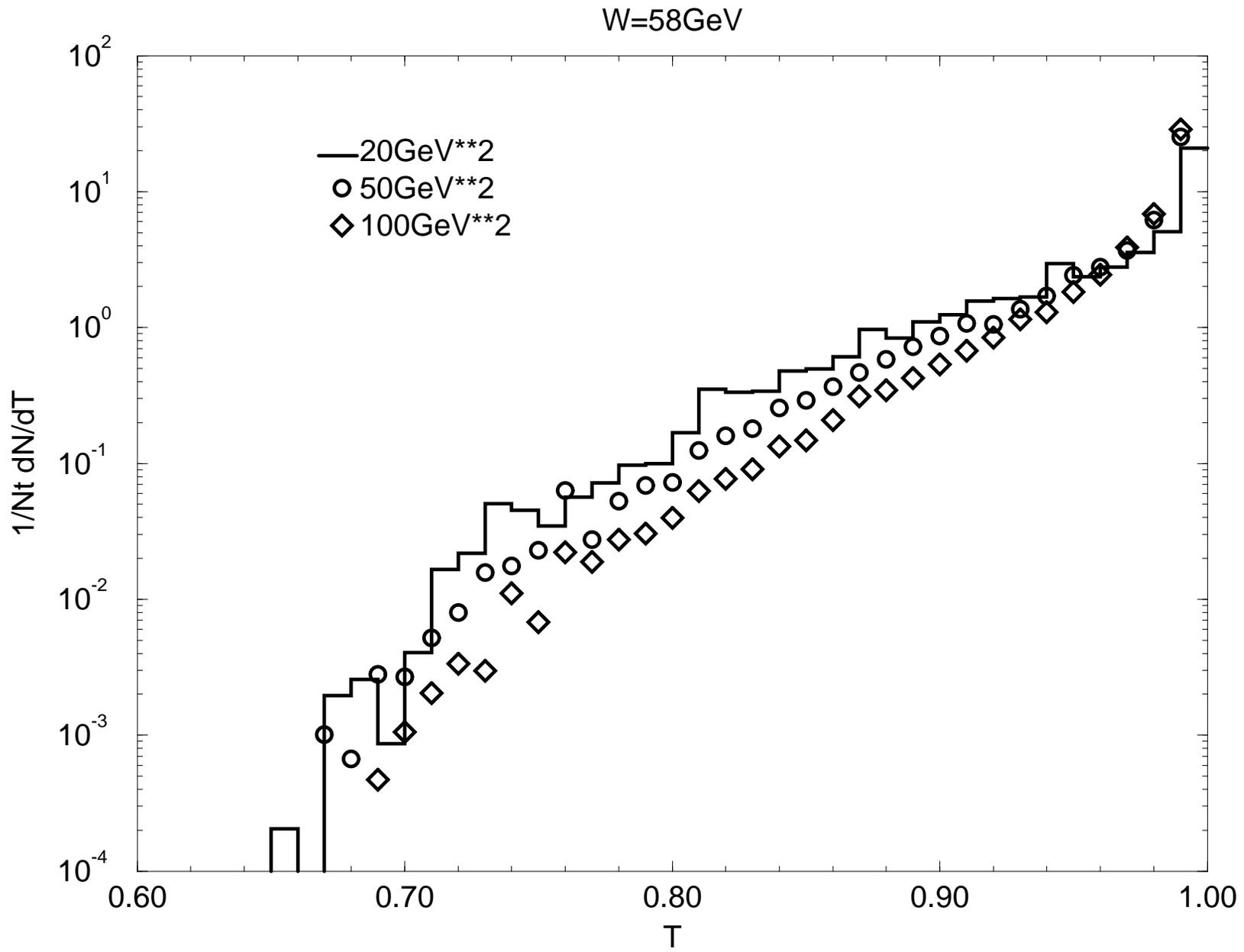
\end{document}